\documentclass[useAMS,usenatbib]{mn2e}

\usepackage{graphics}

\title[200\,Mpc Structure in the 2QZ Survey]
{200\,Mpc Sized Structure in the 2dF QSO Redshift Survey}
\author[L. Miller et al.]
{L. Miller$^{1}$,\thanks{http://www-astro.physics.ox.ac.uk/${\tiny \sim}$lam} 
S.M. Croom$^{2}$,
B.J. Boyle$^{3}$,
N.S. Loaring$^{4}$,
R.J. Smith$^{5}$,
T. Shanks$^{6}$,\newauthor
P. Outram$^{6}$.
\\
$^{1}$Department of Physics, Oxford University,
Denys Wilkinson Building, Keble Road, Oxford OX1 3RH, U.K.\\
$^{2}$Anglo-Australian Observatory, PO Box 296, Epping, NSW 2121, Australia.\\
$^{3}$Australia Telescope National Facility, PO Box 76, Epping, NSW 1710, Australia.\\
$^{4}$Mullard Space Science Laboratory, University College London,
Holmbury St. Mary, Dorking, Surrey RH5 6NT, U.K.\\
$^{5}$Astrophysics Research Institute, Liverpool John Moores University, 
Twelve Quays House, Egerton Wharf, Birkenhead CH41 1LD, U.K.\\
$^{6}$Department of Physics, Durham University,
Science Laboratories, South Road, Durham DH1 3LE, U.K.
}

\begin{document}

\date{Accepted 2004 August 17. Received 2004 July 28: in original form 2004 March 1}

\volume{355} \pagerange{385--394} \pubyear{2004}

\maketitle

\label{firstpage}

\begin{abstract}
The completed 2dF QSO Redshift (2QZ) Survey has been used to search for extreme
large-scale cosmological structure ($\sim$200\,h$^{-1}$\,Mpc) over the
redshift range $0 < z < 2.5$.  We demonstrate that statistically significant
overdensities and underdensities do exist and hence represent the
detection of cosmological fluctuations on comoving scales that correspond to those
presently detected in the cosmic microwave background.  However, the fractional
overdensities on scales $>100$\,h$^{-1}$\,Mpc are in the linear or only weakly 
non-linear regime and do
not represent collapsed non-linear structures.  We compare the measurements
with the expectation of the $\Lambda$CDM model by measuring the
variance of counts in cells and find that, provided
the distribution of QSOs on large scales exhibits a mild bias with respect to
the distribution of dark matter, the observed fluctuations are found to be in good
agreement with the model.  There is no evidence on such scales for any extreme
structures that might require, for example, departures from the assumption of
Gaussian initial perturbations.  Thus the power-spectrum derived from the 
2QZ Survey appears to provide a complete description of the distribution of QSOs.
The amount of bias and its redshift
dependence that is required is consistent with that found from studying the
clustering of 2QZ QSOs on $\sim$10\,h$^{-1}$\,Mpc scales, and may be adequately
described by an approximately redshift-invariant power spectrum with 
normalisation $\sigma_8 \simeq 1.0$ corresponding to a bias at $z=0$ of
$b \simeq 1.1$ rising to $b \simeq 2$ at the survey's mean redshift $z \simeq 1.5$.
\end{abstract}


\section{Introduction}
It has long been recognised that active galaxies and QSOs may used to
probe structure in the universe on the largest observable
scales, and statistical analysis of the clustering of samples of such
cosmological objects was first developed thirty years ago
(e.g. \citealt{webster76,seldner78}).  Even in those early studies it
became apparent that, although the universe appeared generally
homogeneous on the largest scales, individual clusters of active galaxies
could be identified \citep{webster82}.  In modern ideas of active
galaxies and of their link to the process of galaxy formation, it is
now thought that luminous QSOs lie at the heart of the most massive
elliptical galaxies (\citealt{kukula01,dunlop03,floyd04}) 
and that, provided the
clustering bias expected for such a population is accounted for 
\citep{sheth99}, studying the clustering of active galaxies is a means
of testing the predictions of hierarchical galaxy formation theory both
on the largest cosmological scales and as a function of cosmic epoch.

Whether non-uniform distributions of QSOs on large scales could in
fact be detected with a high degree of statistical confidence was an
open question for many years, however, largely because of the limited
sample sizes and systematic uncertainties and varying selection biases
present in samples of QSOs.  A statistically significant detection of
QSO clustering on $\sim$10\,Mpc scales was achieved primarily
from the AAT QSO redshift survey \citep{shanks94}.  With the advent of the
much larger 2dF QSO Redshift Survey (2QZ: \citealt{croom01a,smith04,croom04}) 
not
only has an extremely secure detection of QSO clustering been achieved,
but also the shape of the correlation function, its bias with
respect to the underlying dark matter distribution, the cosmic evolution
in that bias, and the luminosity dependence of QSO clustering have now
all been measured on scales out to about 80\,Mpc
(\citealt{croom01b,croom02,croom03,croom04b}).  

On larger scales, the shape of the power spectrum
of QSO clustering has been measured and shown to be consistent with  
the expectations for a low density CDM universe \citep{outram03}.
The aim of this paper is to look at the distribution of QSOs in 
redshift space, rather than Fourier; to investigate on what scales
departures from randomness may be detected; and 
to test whether those departures
are consistent with the expectation of the widely-assumed cosmological
model in which structures have grown by gravitational collapse of
initially Gaussian fluctuations in a universe dominated by a combination
of cold dark matter and dark energy.  Over the years there have
been a number of claims of the detection of extremely large (50--200\,Mpc)
groupings of AGN and QSOs \citep{crampton87,clowes91,williger02,brand03}:
at the time of these claims there was no consensus on the cosmological
model or on the relationship of the distribution of QSOs to the distribution
of matter, and it was hoped that the measurement of such large groupings might
constrain models of one or both of these.  Now, however, there have emerged 
in the literature preferences for one particular cosmological model \citep{spergel03}
and for the bias of QSOs measured by the 2dF QSO Redshift Survey
\citep{croom01b,croom02,croom03,croom04b}, 
so it is timely to ask whether the very large scale 
distribution of QSOs is consistent with these recent developments.  
If such large-scale clusters
were shown to be collapsed structures with high values of overdensity
this would be likely to be in significant disagreement with the cosmological
model and measurements of QSO bias.  
In fact, we shall see in this paper that indeed very large
groupings of QSOs may be identified, possibly on scales as large as 300\,Mpc,
but that the groupings are still in the linear regime of gravitational
collapse and are in fact entirely in accord with current expectations.

\section{Analysis of the 2QZ survey}
\subsection{The 2QZ survey}
The 2QZ survey is described by \citet{croom01a, smith04, croom04}.  It comprises spectroscopic 
observations of blue colour-selected QSOs in two sky areas totalling 
740\,deg$^2$:  one a strip 5 degrees in declination by 75 degrees in
right ascension passing through the south Galactic pole (the ``SGP'' region)
and the other a similar-sized strip on the celestial equator in the region
of the north Galactic cap (the ``NGP'' region).  In this paper
we analyse each region separately in order to test for systematic errors
or differences between the two halves.  Spectroscopic observations at
$b_J > 18.25$ were carried out with the 2dF facility at the Anglo-Australian
telescope, brighter candidates were observed with the 6dF facility at
the U.K. Schmidt Telescope operated by the Anglo-Australian Observatory but
are not included in this analysis.  
The survey contains 21,181 ``Quality 1'' QSOs with $0 < z < 2.5$ and with photographic
blue magnitude in the range $18.25 < b_J < 20.85$, 
and this is the sample used in the following analysis. 

\subsection{The 2QZ selection function and extinction correction}
In order to carry out the analysis in this paper it is essential to have
an accurate measure of the survey's selection function.  This question has
been extensively discussed by \citet{croom04} and here we adopt the 
same empirical approach to determining the selection function.  

\begin{figure}
\resizebox{75mm}{!}{
\rotatebox{-90}{
\includegraphics{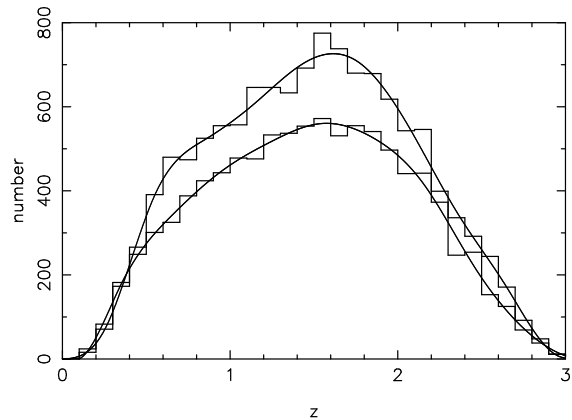}
}}
\caption{
The redshift distributions of QSOs in the 2QZ survey, binned in redshift
intervals of 0.1: (upper) SGP region;
(lower) NGP region.  The smooth curves are the cubic spline fits to the 
observed distributions.}
\label{fig:zdist}
\end{figure}

To do this, we shall
assume that the redshift and angular selection functions are statistically
independent and hence may be separated - we shall discuss later the validity of
this assumption.  The redshift selection function is determined simply
by fitting a cubic spline to the observed redshift distributions
in each of the NGP and SGP regions, in the range $0 < z < 3$.  The cubic
spline was fitted to the data binned in redshift intervals of 0.1 with knots
at redshift intervals of 0.3.   Histograms of redshift and
the cubic spline fits are shown in Fig\,\ref{fig:zdist}. The choice of smoothing
that results from the fitted function may in principle suppress the signal
arising from real clustering on the largest scales, 
but by averaging
over all fields the effects of QSO clustering on the redshift
distribution should be averaged out, and the resulting function should be a 
good reflection of the true selection function.

It may be seen that there
are some small differences between the observed redshift distributions
of the NGP and SGP regions which may reflect some differences in QSO
candidate selection or which may reflect the presence of 
large-scale structure.  In this analysis we analyse each region independently,
using the smoothed redshift distribution from each, in order to be able
to compare the results from the two independent regions and hence
to search for any otherwise-hidden systematic effects.

The photometric colour selection used to construct the survey becomes
inefficient at $z > 2.5$ \citep{croom04} and, as we are concerned about 
fluctuations in selection efficiency possibly mimicking genuine
cosmological structure, we have chosen to truncate the analysis at this
maximum redshift.

The angular selection function is determined by the distribution of
photographic plates used for the initial candidate selection, by the magnitude limit 
and any associated calibration errors, the placement of each spectroscopic 2dF
field and the configuration of observed targets within each field, and the
spectroscopic completeness obtained in each field.  Inspection of the variation
in QSO numbers between the photographic plates gives no indication of any
significant calibration errors which would affect the results presented here.
We shall return to this question below when discussing the results.

To determine the configuration completeness 
at any location on the sky, we again follow \citet{croom04}. 
Each spectroscopic observation was carried out in a 2dF field of diameter
two degrees.  These circular regions were overlapped in order to provide
contiguous coverage of the sky, and the intervals between field centres
were chosen to optimise the fibre acquisition of targets.  Hence we may
define a set of sectors defined by the boundaries of the 2dF fields:  a
sector may contain data from a single 2dF observation or from overlapping
2dF observations, depending on the location.  The survey comprises
3751 such sectors.  Within each sector, we count the fraction
of the colour-selected targets that were observed spectroscopically, and
define that as the configuration completeness at that location.  
227 regions around bright stars were removed from the survey.
87\,percent of the remaining survey area has configuration 
completeness higher than 90\,percent. 

However, such a measure does not take into account that the efficiency of
identifying QSOs may vary between 2dF fields, depending on observing
conditions.  
Hence an alternative completeness measure is the spectroscopic
completeness, also discussed in detail by \citet{croom04}.  In each sector we
determine the fraction of all colour-selected targets that have been
spectroscopically identified.  Hence this measure includes the configuration
completeness, but allows additional variation in spectral quality.

Although it might seem better to use the spectroscopic completeness rather
than just the configuration completeness, there is a concern that this
could bias the results, because of a combination of two effects.  First, it
is generally easier to identify QSOs than stars from their spectra, because
of the presence of strong broad emission lines in the former.  Second, the
fraction of targets that actually are QSOs varies with sky position,
because the surface density of stars varies with Galactic latitude and
longitude.  Hence regions of high star density will tend to have lower
spectroscopic completeness, even though in fact the efficiency of QSO
detection may be invariant.  This completeness measure is also sensitive to
variations in the colour selection with position:  regions with colour 
selection that is redder, owing to photometric calibration uncertainty,
would have a higher fraction of stellar contaminants, and hence the spectroscopic
completeness would appear lower, whereas again the QSO completeness may
be invariant.

In practice, variations in spectroscopic
completeness between 2dF fields are unlikely to degrade significantly the
results presented below, because the angular extent of the 2dF 
fields is smaller than the physical scales of interest over most of the redshift
range, and because neighbouring 2dF fields were not typically observed 
contemporaneously.  Hence the spectroscopic completeness variations 
should to some extent average out on the scales of interest.  For this reason 
we prefer to use the configuration completeness as the baseline measure, as
this is less susceptible to the other problems discussed above, but in the 
quantitative analysis that follows we compare results obtained assuming both
completeness measures, as an indication of the possible magnitude of residual
systematic uncertainties arising from uncertainties in the observational
selection. 

Before leaving the discussion of completeness corrections, we should
note that any effects of incorrect completeness
corrections or significant photometric calibration variations
should be observable as a characteristic signature on the maps
of fluctuations presented below.  A field containing more QSOs than its
neighbours owing to a completeness variation would reveal itself as a radial
feature in those maps.  No such features are visible at the level of fluctuations
detected.

The final ingredient which should be included as part of the selection function
is the effect of Galactic extinction.  A region of high extinction effectively
makes the QSO magnitude selection limit brighter, reducing numbers of observed
QSOs in that region.  We correct for the known Galactic extinction using the
dust maps of \citet{schlegel} with resolution of 6\,arcmin,
an adequate resolution for the detection of large-scale fluctuations in
QSO numbers.  The catalogued values of $E(B-V)$ are converted to extinction
assuming $A_B = 4.315 E(B-V)$ and the effect on QSO numbers is calculated
assuming a slope for the QSO number-magnitude relation at the faint magnitude
limit of 0.32.  Extinction in the NGP region is significantly higher than in
the SGP region, yet, after application of the extinction correction, no
statistically significant systematic difference remains in the results from
the two regions.

In the analysis presented below the observed fluctuations in QSO numbers
are estimated from a comparison of the actual QSO distribution with 
artificial distributions created from numerical realisations.
The completeness and extinction corrections are applied to those realisations
in order to obtain simulated datasets that should mimic the true data to
high accuracy.

\subsection{The detection of QSO density fluctuations}
A number of proposed methods of detecting clusters of objects are
described in the literature, but perhaps the most straight-forward
real-space (or redshift-space) measure 
in terms of understanding its statistics is to
smooth the 3D distribution of QSOs and to search for statistically
significant departures from a uniform distribution, taking into account
sample selection and 
the shot noise arising from the finite sample size.  If significant
fluctuations on some scale are detected then we may further compare
the results with the expectations of the $\Lambda$CDM cosmological model:
this analysis is carried out in the next section.  In the following
analysis we assume that redshift-space distortions may be neglected 
on the scales of interest at the amplitudes being investigated.  

\begin{figure*}
\resizebox{125mm}{!}{
\rotatebox{-90}{
\includegraphics{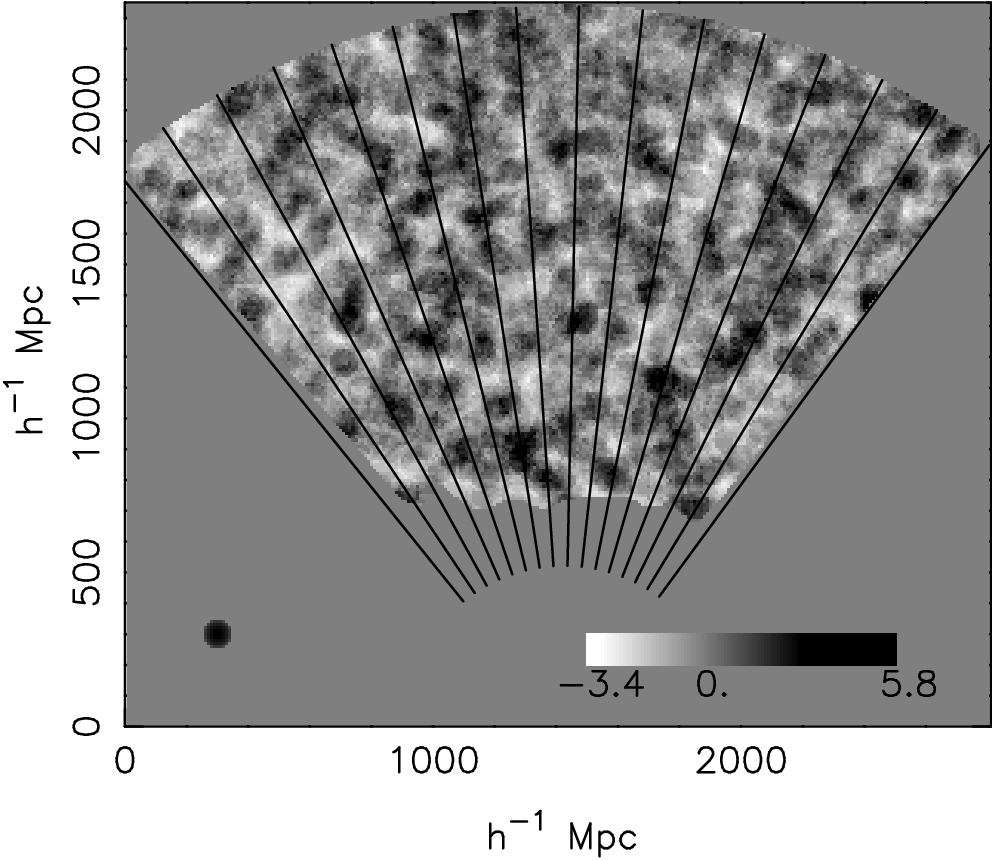}
}}
\resizebox{125mm}{!}{
\rotatebox{-90}{
\includegraphics{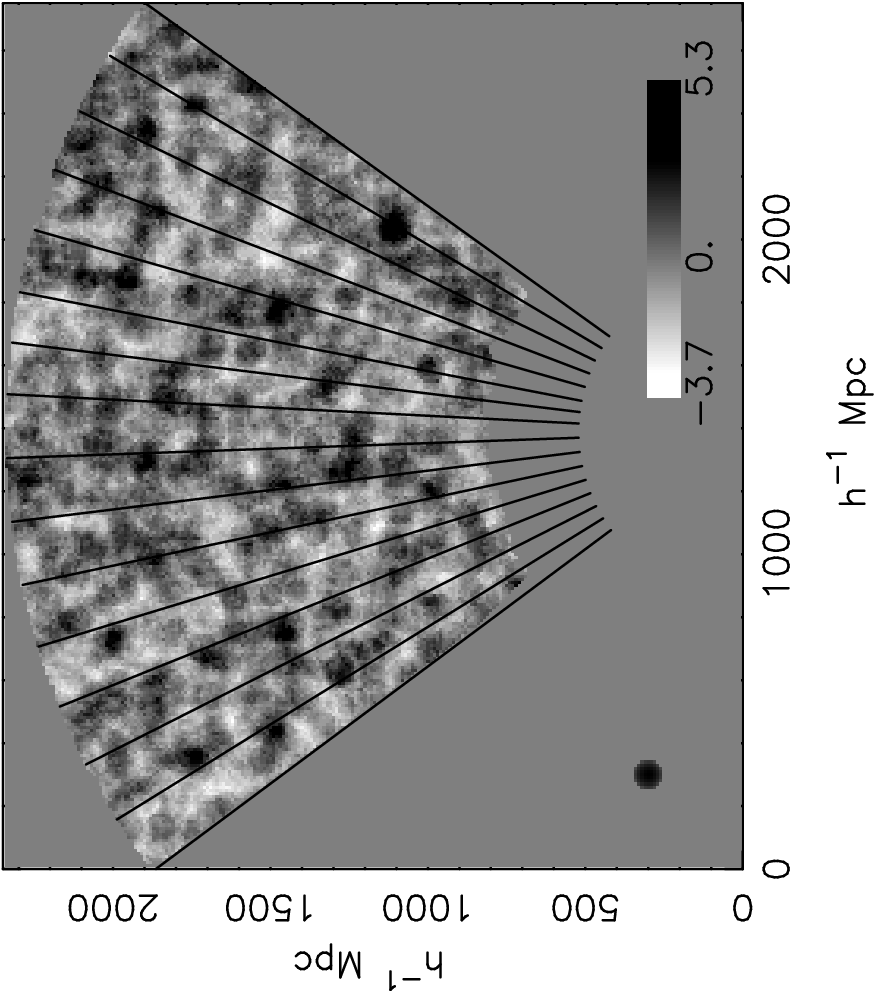}
}}
\caption{
Cuts at constant declination through the smoothed three-dimensional
maps of fluctuations
in the NGP (upper) and SGP (lower) regions.  The x-axis corresponds
crudely to the right ascension direction, the y-axis to the redshift
direction.  The smoothing function is a
spherical top hat of diameter 100\,h$^{-1}$\,Mpc as indicated on the lower
left of each map.  The grey scale is in units of signal-to-noise ratio $S$, 
as defined in the text, 
and is the same on both plots.  The greyscale bar indicates
the maximum and minimum values found in each region.  
Also shown are the boundaries defined by the right ascension
limits of the UKST photographic plates used to construct the survey.
\label{fig:SN50}
}
\end{figure*}

\begin{figure*}
\resizebox{125mm}{!}{
\rotatebox{-90}{
\includegraphics{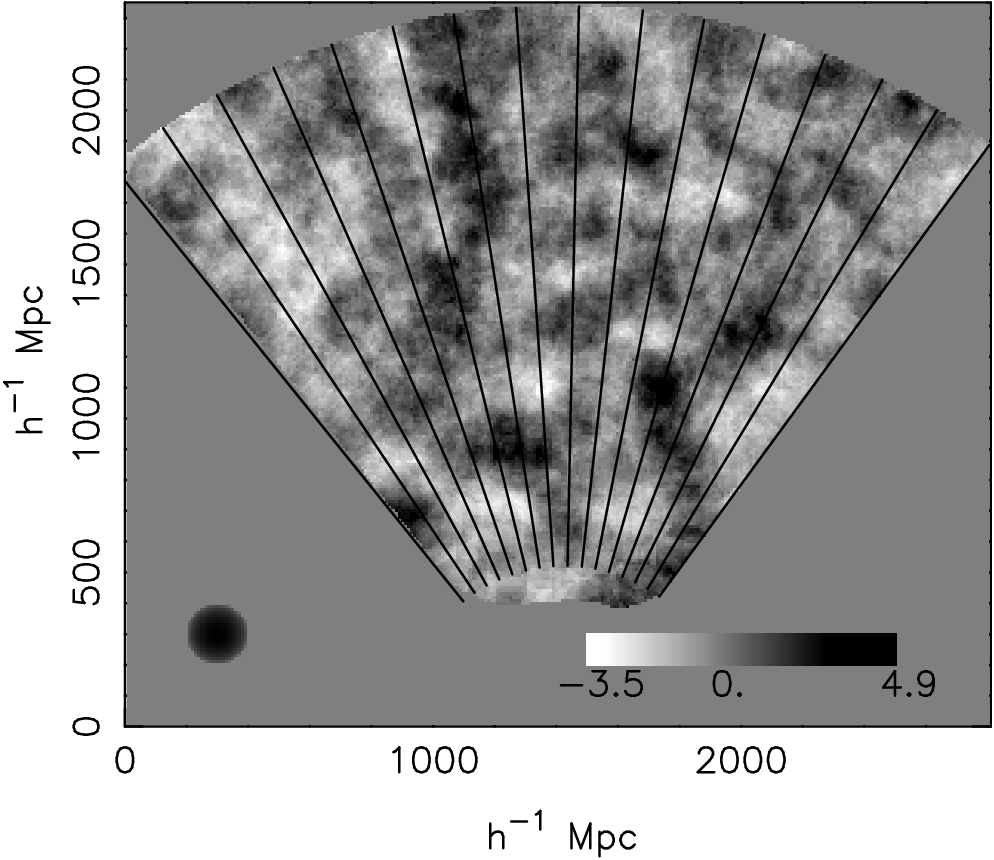}
}}
\resizebox{125mm}{!}{
\rotatebox{-90}{
\includegraphics{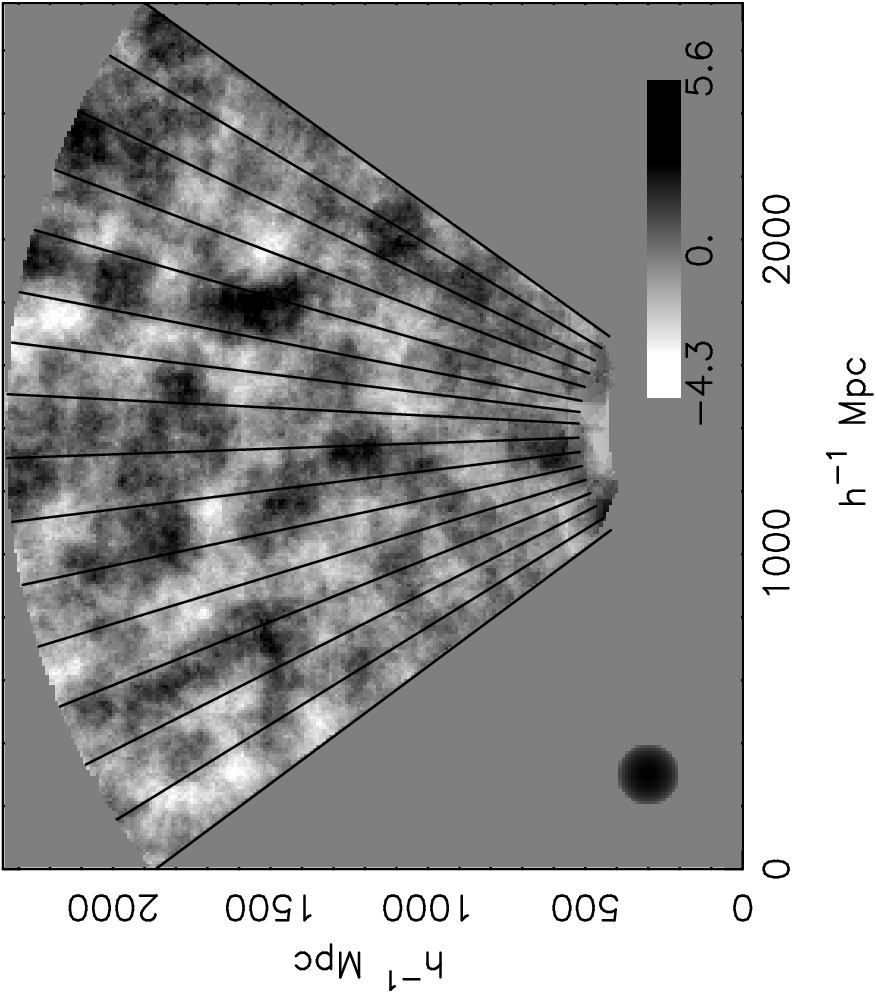}
}}
\caption{
As Fig.\ref{fig:SN50}, but with a 
spherical top hat smoothing function of diameter 200\,h$^{-1}$\,Mpc as indicated 
on the lower left of each map.  
\label{fig:SN}
}
\end{figure*}

As we are interested in trying to detect structures on the largest scales that
have previously been claimed, we smooth the QSO distribution with a 
spherical top-hat function of diameter 100 or 200\,h$^{-1}$\,Mpc.  The smoothed
density distribution is evaluated at a sampling of 10\,h$^{-1}$\,Mpc.
The locations of QSOs are determined assuming the redshift is purely
Hubble-flow and assuming a $\Lambda$CDM cosmological model with 
$\Omega_M = 0.3$, $\Omega_{\Lambda} = 0.7$.
In order to compare the observed distribution with the distribution expected
in the absence of real density fluctuations, we also compute the
numbers expected given the survey selection function, smoothed and sampled
the same as the data.  Figs\,\ref{fig:SN50}\,\&\,\ref{fig:SN} plot cuts at constant declination
through each of the two halves of the 2QZ survey for two choices of smoothing
scale.  The plotted scale is
the signal-to-noise ratio of the observed variations in QSO numbers, estimated 
as
$$
S = \frac{n_{\rm obs} - n_{\rm exp}}{\sqrt n_{\rm exp}}.
$$
where $n_{\rm obs}$ is the observed number of QSOs at a given location in the smoothed
distribution and $ n_{\rm exp}$ is the expected number given the selection
function and in the absence of any cosmological structure.  Only regions with
$n_{\rm exp} > 10$ are plotted.
The statistic $S$ may only be interpreted as having a 
normal distribution in the limit of large values of $n_{\rm exp}$, but
in fact given the typical value of $n_{\rm exp} \sim 100$ this is a reasonable
approximation, and leads to the observation that when smoothed on this scale,
there exist statistically significant deviations from the null
hypothesis that QSOs trace a randomly sampled uniform distribution.  
The regions that appear significantly under- or over-dense are resolved
on these maps, and it appears that significant excesses may be traced out to
substantially larger scales than the 200\,h$^{-1}$\,Mpc smoothing scale.
Although the top-hat filter has undesirable properties when transformed to Fourier
space, in the analysis presented here one can be confident that distant points
in the smoothed distribution have uncorrelated shot noise, and the amplitude of
fluctuations observed in the smoothed maps may be more readily compared with
the analysis of the variance of counts in cells that is presented in the next section.

We should perhaps be concerned that errors or uncertainties in the 2QZ
selection function may result in spurious creation of fluctuations in 
Figs\,\ref{fig:SN50}\,\&\,\ref{fig:SN}.
The strongest argument against this possibility are that the selection
function and extinction correction have been entirely separated into
radial and angular components, and we would expect errors in either
component to be manifest as purely radial or angular signals.  
In Figs\,\ref{fig:SN50}\,\&\,\ref{fig:SN} 
we also show the boundaries defined by the right ascension
limits of the UKST photographic plates used to construct the survey.  There is
no obvious tendency for there to be radial overdensities aligned with these
boundaries:  although the alignment of a minority of the features appears radial,
this is not generally the case, and those features that could be radial in nature
may be seen to not fill an entire UKST photographic plate and to extend across
the plate boundaries.  Overall, we conclude that there are no obvious radial 
signals associated with the photographic plates.
At a greater level of subtlety, there may be variations
in the selection function which are not well represented by the assumption
of separable components (i.e. angular-dependent variations in the redshift
selection function).  However, if such effects are present, the existence
of under- and over-dense regions apparently at randomly placed redshifts,
and not particularly restricted to individual photographic plates, 
implies that such an effect is not dominating the observed distributions.

\begin{figure*}
\resizebox{125mm}{!}{
\rotatebox{-90}{
\includegraphics{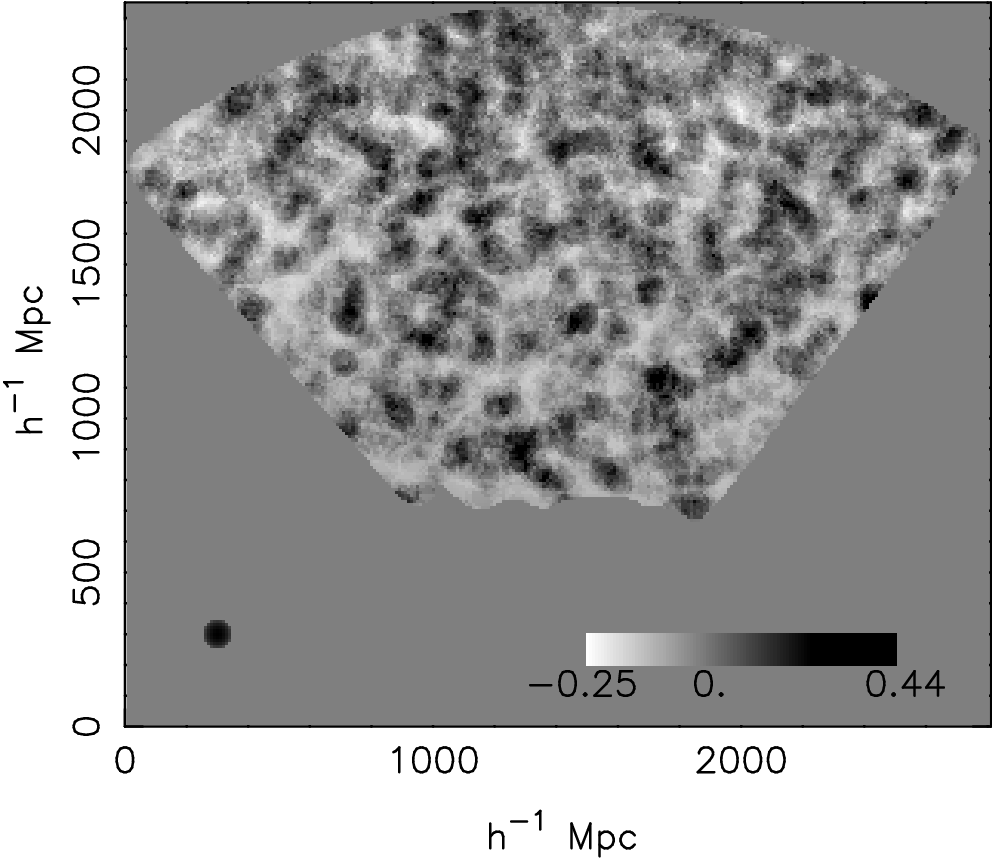}
}}
\resizebox{125mm}{!}{
\rotatebox{-90}{
\includegraphics{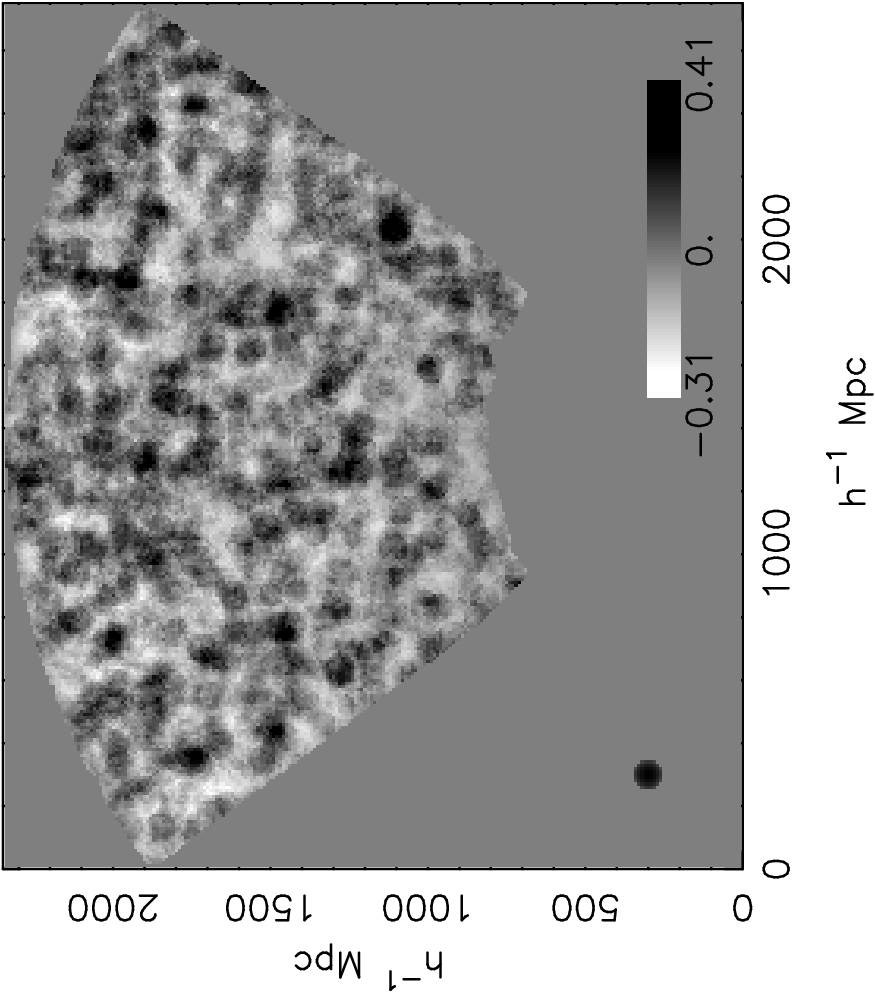}
}}
\caption{
Cuts at constant declination through the Bayesian maps of smoothed fluctuations
in the NGP (upper) and SGP (lower) regions.  The smoothing function is a
spherical top hat of diameter 100\,h$^{-1}$\,Mpc as indicated on the lower
left of each map.  The grey scale is in units of 
fractional overdensity and is the same on both plots.  The bar also indicates
the maximum and minimum values found in each region.
\label{fig:Bayesianmap50}
}
\end{figure*}

\begin{figure*}
\resizebox{125mm}{!}{
\rotatebox{-90}{
\includegraphics{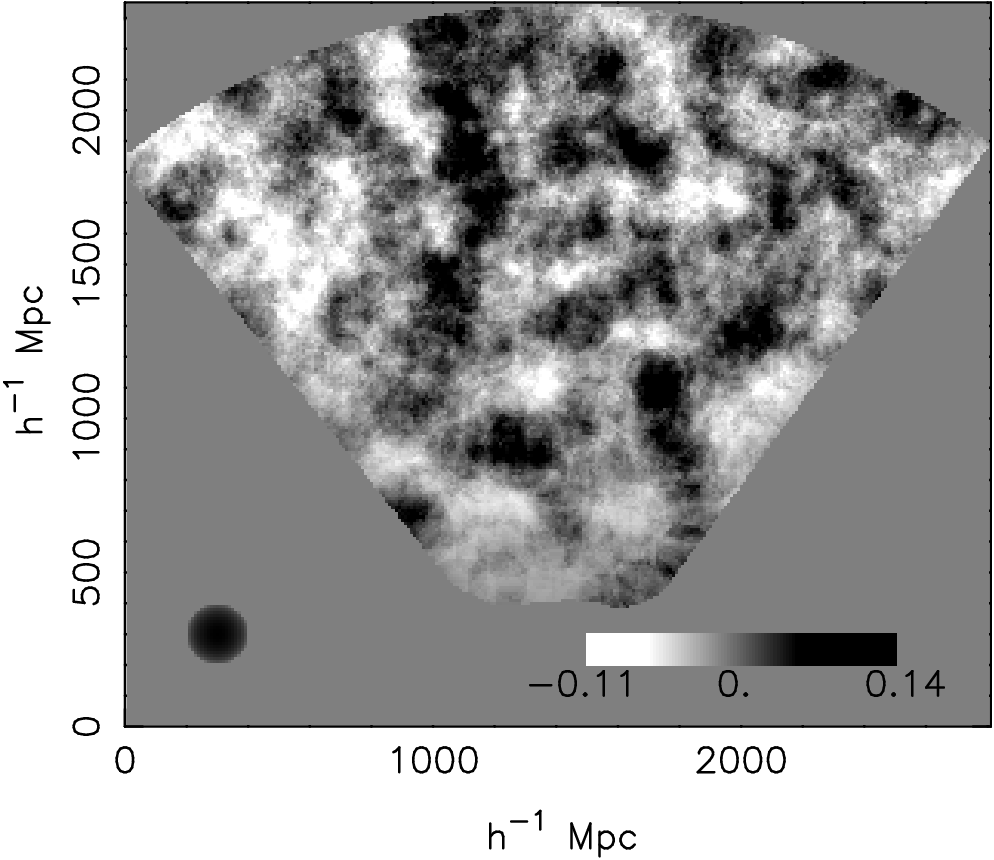}
}}
\resizebox{125mm}{!}{
\rotatebox{-90}{
\includegraphics{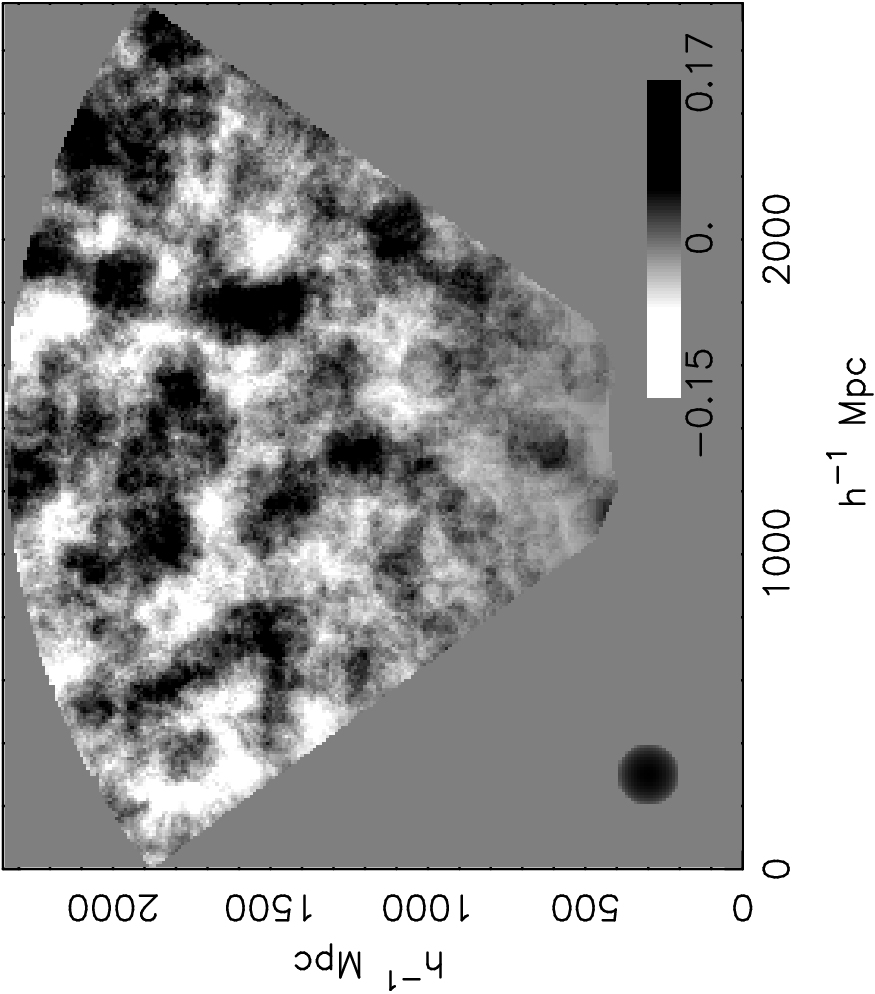}
}}
\caption{
As Fig.\ref{fig:Bayesianmap50} but with a 
spherical top hat smoothing function of diameter 200\,h$^{-1}$\,Mpc.  
\label{fig:Bayesianmap}
}
\end{figure*}

Figs\,\ref{fig:Bayesianmap50}\,\&\,\ref{fig:Bayesianmap} show the same data 
but now presented as plots of fractional overdensity
in QSO number.  Because shot noise cannot be neglected, for this exercise 
we adopt a Bayesian estimator of the 
fractional overdensity at each location on the map, which maximises the
posterior probability 
$$
P\left(\delta | n_{\rm obs},n_{\rm exp},\sigma^2 \right) \propto 
P\left(\delta | \sigma^2 \right) 
P\left(n_{\rm obs} | \delta, n_{\rm exp} \right)
$$
where $\delta$ is the fractional overdensity and 
$\sigma^2$ is the cosmological variance in $\delta$ on the scale being considered.
For Gaussian fluctuations 
$$
P\left(\delta | \sigma^2 \right) = \frac{1}{\sqrt{2\pi\sigma^2} }\exp{\frac{-\delta^2}{2\sigma^2}}.
$$
The Bayesian estimate has the property that in the presence of random noise
the expectation value of a set of points with the same estimated value
is the true value, and in that sense it provides the ``most likely'' value
of fluctuation at any point on the map.
A maximum likelihood estimate would have the
property that the expectation value of a set of points with the same 
true value is that value.
Taking the maximum likelihood estimate in the presence of random
noise leads to the statistical
distribution of fluctuations being broader than the true distribution,
whereas the Bayesian estimator leads to a distribution that is narrower
than the true distribution.  On the maps presented here, regions of low
signal-to-noise have Bayesian estimates of overdensity that are biased towards
zero (if a maximum likelihood estimate had been applied instead, regions of
low signal-to-noise would appear biased to values more extreme than reality).
This effect is noticeable in 
Figs\,\ref{fig:Bayesianmap50}\,\&\,\ref{fig:Bayesianmap}
at low redshift, where the increasing shot noise leads to some dilution
of the measured signal.
The Bayesian estimates were created assuming the $\Lambda$CDM model 
of Section\,2.4 with $\sigma_8 = 1.0$.  

It is immediately seen that the detected fluctuations are all
in the linear or weakly non-linear regime, with 
$-0.31 < \delta < 0.44$ on scales of 100\,h$^{-1}$\,Mpc and 
$-0.13 < \delta < 0.15$ on scales of 200\,h$^{-1}$\,Mpc.  
Hence the 
chief result of this section is that large-scale fluctuations in QSO numbers
exist, as suspected by previous authors, but unless the bias is much less than
unity (the issue of QSO bias is discussed later) the mass density fluctuations
are not of sufficiently large amplitude to represent collapsed structures.  Hence the
best matching hypothesis is that the fluctuations are simply those that are
expected in the linear regime of the growth of cosmological structure, and 
that these are fully represented statistically by the power spectrum of the
2QZ survey \citep{outram03}.  This hypothesis is tested further in the next section.

\subsection{A counts-in-cells analysis}
Having demonstrated that statistically-significant fluctuations do exist on large 
scales, we now should test how these fluctuations compare with the predictions
of the cosmological model assuming that on large scales structures are
fully represented by the power spectrum.  To do this we carry out a counts-in-cells
analysis, measuring the sample variance when the data are sampled in independent
cubes of a specified size, and comparing that sample variance with the variance
expected given a particular choice of power spectrum and normalisation.

The measured cell-to-cell variance has contributions from two sources:  cosmological
fluctuations and shot noise.  Furthermore, because the QSO density varies over the
survey, the weight given to each cell should be varied to optimise the signal-to-noise
of the overall sample variance.  In the limit where QSO numbers are sufficiently large
that the shot-noise Poisson distribution tends to a normal distribution, the
maximum likelihood estimate of variance of overdensity $\delta$ is
$$
s^2 = \frac{\sum \left(n_{\rm obs}-n_{\rm exp} \right)^2 - \sum n_{\rm exp}}
{\sum n_{\rm exp}^2}.
$$
As here $n_{\rm exp} >> 1$ we adopt this statistic.  However,
as comparison is made with values of this statistic measured in simulated data
(see below) we do not need to assume that the central limit theorem holds.
The statistic $s^2$ is calculated from the data for cell sizes of 
80, 160, 320 \& 640\,h$^{-1}$\,Mpc.  The use of cubic cells means that,
for a given cell size, each cell
has shot noise that is statistically independent from its neighbour (although
the cosmological signal does not have this property).  The variances we expect
then are slightly different from the variances that would be calculated using
a more standard assumption of spherical cells.

In order to calculate the expected variance for a given cosmological model, we could
in principle carry out the standard integration over the power spectrum
(e.g. \citealt{peacock}) assuming an appropriate cell size and shape.  
There are a number of reasons not to do this, however.  First, the surveyed volume
does not fill a geometrically simple region, and in fact on the largest scales
probed here the survey becomes two-dimensional at low redshifts:  the survey
dimension in the declination direction is smaller than the largest cell size.
This causes the expected variance to have a modified dependence on cell size.
Second, the fluctuations in cells of differing sizes are correlated, because the
same data are used in the analysis of different cell sizes.  The degree of
correlation depends on the selection function, the power spectrum, 
the number of QSOs and the choice of statistic used to measure the variance.  

Hence the approach adopted here is
to create mock QSO catalogues which have the same selection function and shot
noise as the actual data, and in which the expected level
of cosmological fluctuations are present, assuming a particular cosmological model.
By averaging the values obtained from a large number of such simulations this
approach automatically includes the effects of cosmic variance in the distribution
of the statistic $s^2$.  The simulations are generated by creating random density fields
on a grid of sampling 10\,h$^{-1}$\,Mpc and size much larger than each region of the
2QZ survey, whose Fourier components are drawn randomly from a normal distribution
with variance specified by the power spectrum.  The density fields are then multiplied by the
2QZ selection function and extinction corrections. These are Poisson-sampled to create 
simulated datasets whose properties should mimic in every way the properties of the actual data.
This process is repeated for the NGP and SGP regions of the survey.

\begin{figure*}
\resizebox{168mm}{!}{
\rotatebox{-90}{
\includegraphics{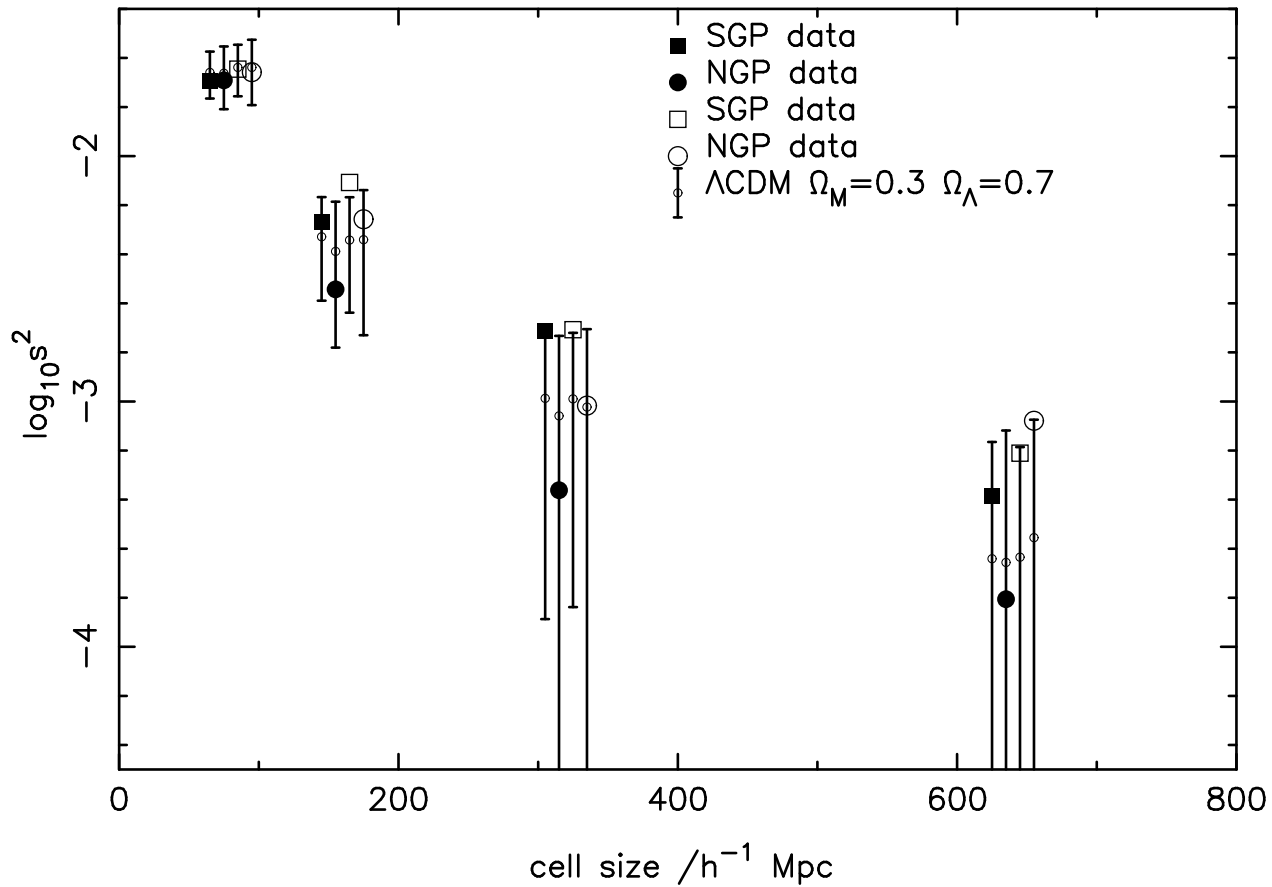}
}}
\caption{
The variance of counts in cells for the NGP and SGP halves of the survey.  Filled points
show the measured results after correcting for configuration completeness.  Open
points show the results after correcting also for spectroscopic completeness.
Also shown are mean values and 68 percent confidence intervals derived from the
numerical simulations for each region and for each of the two completeness corrections.
A value of $\sigma_8 = 1.0$, which best fits the filled points, has been adopted.
\label{fig:variance}}
\end{figure*}

The shape of power spectrum adopted is that of \citet{efstathiou92}.  
We assume a $\Lambda$CDM cosmology with $\Omega_{\Lambda}=0.7$, $\Omega_{M}=0.3$,
$H_0=70$\,km\,s$^{-1}$\,Mpc$^{-1}$.  To set the normalisation we need
to consider the bias of the power spectrum, but dealing with this is complicated by
the possibility that departures of the Poisson shot noise from a normal distribution
could lead to an offset between the values of the statistic $s^2$ measured for the
data and for the simulations.  Hence we have tested a range of values of the
normalisation parameter $\sigma_8$ (the rms fluctuation in spheres of diameter
8\,h$^{-1}$\,Mpc), and choose the value which produces the best fit of the
statistic $s^2$ to the data.  The dark matter power spectrum also evolves in 
amplitude with redshift:  what should we assume for the power spectrum of QSO
number fluctuations?  In fact, we expect QSOs to exhibit larger fluctuations than
the dark matter, because QSOs exist in the most massive individual galaxies
known \citep{dunlop03}, and hence the amplitude of QSO fluctuations will be biased,
as described by, e.g. \citet{sheth99}.  Because massive galaxies are increasingly
rare at higher redshifts, the bias increases with redshift.  Hence the expected
amplitude of QSO fluctuations should have a redshift dependence which is dependent
on the mass of QSO host galaxies.  Rather than assuming a theoretical value for
the evolution of the bias, we can measure it from the 2QZ survey on smaller physical
scales (\citealt{croom01b,croom03,croom04b}, Loaring et al. in prep.) and assume that the
cosmological evolution of the bias is the same on large scales as on 10\,Mpc scales.
To a good approximation, in a $\Lambda$CDM cosmology, the 10\,Mpc-scale clustering
has an amplitude which is constant with redshift over the range $0.5 < z < 2$, so
here we shall assume that the power-spectrum amplitude is invariant with redshift. 
The uncertainties in the variance of the cell counts are large,
so it should make little difference if in reality there is some departure from 
the ``no-evolution'' assumption (in fact, it would be  more correct to
say that there is strong evolution of the bias which
cancels the evolution in the dark matter fluctuations).

Fig.\,\ref{fig:variance} 
shows the measured variance as a function of cell size for the two halves of the survey.
It may be seen that the two halves are in good agreement.  
Values are shown for both methods of completeness correction: configuration (filled symbols)
and spectroscopic (open symbols).  Also shown on the figure 
is the expected variance from the mean of 100 $\Lambda$CDM simulations for a
QSO power spectrum normalisation of $\sigma_8=1.0$, this being the best fitting 
redshift-independent power spectrum normalisation.  Extrapolated to $z=0$ this would
correspond to a bias value $b\simeq 1.1$, but at the typical redshift of the 2QZ sample
the bias is substantially higher, consistent with the findings based on measuring the
correlation function on smaller scales \citep{croom02,croom03,croom04b}.  
The error bars shown reflect
the 68\,percent range of values in cell variance obtained from the simulations and are
therefore good estimates of the random errors, taking into account both shot noise and
cosmic variance.  The total $\chi^2$ of the fit to both regions
is 1.8 with 7 degrees of freedom, assuming the selection function determined from
the configuration completeness. The covariance between data points
in Fig.\,\ref{fig:variance} is low but is included in the $\chi^2$ estimation.
The measured variance increases and the best-fitting  
$\chi^2$ value rises slightly to 5.3 if the spectroscopic
completeness correction is adopted (open symbols in Fig.\,\ref{fig:variance}).
It may be seen that on the largest scales
the two completeness correction methods reveal systematic uncertainties in the variance
that are comparable to the size of the random errors (shot noise and cosmic variance).

If we assume $\sigma_8=0$ (i.e. that there are no cosmological fluctuations)
$\chi^2$ rises to 61, indicating that the hypothesis that the observed fluctuations are
merely due to shot noise may be rejected at a significance level of $3 \times 10^{-10}$.

We note at this stage that we have not attempted to compare the data with any
competing models, but rather simply to test whether the data are consistent with the
model favoured by independent cosmological measurements: the approach adopted 
here may be viewed as being frequentist in nature rather than Bayesian.  However,
we might wonder whether the shapes and topology of the detected structures are
as expected within this model, or whether there is any evidence for significant
differences.  Statistical tests of topology are beyond the scope of this paper,
but by comparing the structures detected on the two scales shown in Figs\,2-5, we can see
qualitatively at least that there is no evidence for, for example, extremely filamentary
structure being responsible for the observed fluctuations on the larger scale.
A qualitative comparison of the appearance of Figs\,4\,\&\,5 with the
$\Lambda$CDM simulated maps (not shown here) also reveals no significant differences.

\section{Discussion and conclusions}
Fluctuations in QSO space density on scales $\sim 200$\,h$^{-1}$\,Mpc may be 
seen directly in
the QSO distribution (Figs\,\ref{fig:SN50}-\ref{fig:Bayesianmap}).  
It is likely that these fluctuations may be traced out to
larger scales, and inspection of the maps and
analysis of the variance of the QSO density field indicates the detection
of fluctuations on scales 
possibly as large as 300\,h$^{-1}$\,Mpc.  The fluctuations are in good
agreement with those expected in a $\Lambda$CDM cosmology 
with WMAP parameters if we assume that QSO
clustering is a biased tracer of dark matter fluctuations, with a bias value that is
approximately redshift independent and which is close to the value inferred from QSO
clustering on 10\,Mpc scales.  Since it appears that $b > 1$, the fluctuations in mass
associated with the observed QSO fluctuations are certainly in the linear or weakly non-linear
regime of gravitational collapse, and hence they do not represent collapsed non-linear
structures, but are merely a reflection of the large-scale fluctuations expected given the
$\Lambda$CDM power spectrum.  Figs\,\ref{fig:SN50}-\ref{fig:Bayesianmap} 
represent direct detection of structure
in the distribution of discrete objects on comoving scales that correspond to the scales
measured in the cosmic microwave background by WMAP \citep{spergel03}.

Consideration of two methods of correcting the measurements for observational selection
shows that on the largest scales 
there may remain systematic variations in survey uniformity at levels that
are comparable to the measured fluctuations.  For this reason we believe that the 2QZ survey
should not be used to attempt to infer more precise values for cosmological
parameters such as $\Omega_M$ until those systematic non-uniformities can be better
corrected for.

One of the most significant features of the results presented here is that there is no
evidence for any non-Gaussian initial conditions, which might have been implied had any of
the $> 50$\,Mpc structures previously claimed for QSO distributions turned out to
represent collapsed structures.
As the counts in cells are completely in accord with the simulations generated from
a $\Lambda$CDM power spectrum, 
we conclude that, as far as this test is concerned, the power spectrum
remains a complete description of the large-scale distribution of QSOs.  

We have not in this paper attempted to measure the topology of the observed fluctuations.
This would be an interesting exercise in order to test whether the observed structures are
consistent in every way with the expectations of linearly collapsing structures forming
from Gaussian initial conditions.  

\vspace*{5mm}

\noindent
{\large \bf ACKNOWLEDGEMENTS}\\

The 2dF QSO Redshift Survey was based on observations made with the U.K.~Schmidt
and Anglo-Australian Telescopes.  We thank all the present and former
staff of the Anglo-Australian Observatory for their work in
building and operating the 2dF facility.

\label{lastpage}


\begin{thebibliography}{99}
\bibitem[\protect\citeauthoryear{Brand et al.}{2003}]{brand03}
Brand, K., Rawlings, S., Hill, G.J., Lacy, M., Mitchell, E. \& Tufts, J.,
2003, MNRAS, 344, 283
\bibitem[\protect\citeauthoryear{Clowes \& Campusano}{1991}]{clowes91}
Clowes, R.G. \& Campusano, L.E., 1991, MNRAS, 249, 218
\bibitem[\protect\citeauthoryear{Crampton et al.}{1987}]{crampton87}
Crampton, D., Cowley, A.P. \& Hartwick, F.D.A., 1987, ApJ, 314, 129
\bibitem[\protect\citeauthoryear{Croom et al.}{2001a}]{croom01a}
Croom, S,M., Smith, R.,J., Boyle, B.J., Shanks, T., Loaring, N.S., Miller, L., 
\& Lewis, I.J., 2001, MNRAS, 322, 29
\bibitem[\protect\citeauthoryear{Croom et al.}{2001b}]{croom01b}
Croom, S,M., Shanks, T., Boyle, B.J., Smith, R.,J., Miller, L., Loaring, N.S.,
\& Hoyle, F., 2001, MNRAS, 325, 483
\bibitem[\protect\citeauthoryear{Croom et al.}{2002}]{croom02}
Croom, S.M., Boyle, B.J., Loaring, N.S., Miller, L., Outram, P.J.; Shanks, T. \& Smith, R.J.,
2002, MNRAS, 335, 459
\bibitem[\protect\citeauthoryear{Croom et al.}{2004a}]{croom03}
Croom, S.M., Boyle, B.J., Shanks, T., Outram, P.J., Smith, R.J.,
Miller, L., Loaring, N.S., Kenyon, S. \& Couch, W., 2004a,
in Richards G.T., Hall, P.B., ASP Conf Series Vol. CS-311,
AGN Physics with the Sloan Digital Sky Survey, Astron. Soc. Pac.,
San Francisco
\bibitem[\protect\citeauthoryear{Croom et al.}{2004b}]{croom04}
Croom, S.M., Smith, R.J., Boyle, B.J., Shanks, T., Miller, L., Outram, P.J. \& Loaring, N.S. 
2004b, MNRAS, 349, 1397
\bibitem[\protect\citeauthoryear{Croom et al.}{2005}]{croom04b}
Croom, S.M., Boyle, B.J., Shanks, T., Smith, R.J., Miller, L., Outram, P.J.,
Loaring, N.S., Hoyle, F. \& Da Angela, J.,
2005, MNRAS, 356, 415
\bibitem[\protect\citeauthoryear{Dunlop et al.}{2003}]{dunlop03}
Dunlop, J.S., McLure, R.J., Kukula, M.J.,
Baum, S.A., O'Dea, C.P. \& Hughes, D.H.,
2003, MNRAS, 340, 1095
\bibitem[\protect\citeauthoryear{Efstathiou et al.}{1992}]{efstathiou92}
Efstathiou, G, Bond. J.R. \& White, S.D.M., 1992, MNRAS, 258, 1P
\bibitem[\protect\citeauthoryear{Floyd et al.}{2003}]{floyd04}
Floyd, D.J.E., Kukula, M.J., Dunlop, J.S., McLure, R.J., Miller, L.,
Percival, W,J., Baum, S,A. \& O'Dea, C.P., 2003,
MNRAS, 355, 196
\bibitem[\protect\citeauthoryear{Kukula et al.}{2001}]{kukula01}
Kukula, M.J., Dunlop, J.S., McLure, R.J., Miller, L.,
Percival, W,J., Baum, S,A. \& O'Dea, C.P., 2001, MNRAS, 326, 1533
\bibitem[\protect\citeauthoryear{Outram et al.}{2003}]{outram03}
Outram, P.J. Hoyle, F., Shanks, T.,
Croom, S.M., Boyle, B.J., Miller, L., Smith, R.J. \& Myers, A.D.,
2003, MNRAS, 342, 483
\bibitem[\protect\citeauthoryear{Peacock}{1999}]{peacock}
Peacock, J.A., ``Cosmological Physics'', 1999, Cambridge University Press
\bibitem[\protect\citeauthoryear{Seldner \& Peebles}{1978}]{seldner78}
Seldner, M. \& Peebles, P.J.E., 1978, ApJ, 225, 7
\bibitem[\protect\citeauthoryear{Schlegel et al.}{1998}]{schlegel}
Schlegel, D., Finkbeiner, D. \& Davis, M.,
ApJ, 1998, 500, 525
\bibitem[\protect\citeauthoryear{Shanks \& Boyle}{1994}]{shanks94}
Shanks, T. \& Boyle, B.J., 1994, MNRAS, 271, 753
\bibitem[\protect\citeauthoryear{Sheth \& Tormen}{1999}]{sheth99}
Sheth, R.K. \& Tormen, G. 1999, MNRAS, 308, 119
\bibitem[\protect\citeauthoryear{Smith et al.}{2004}]{smith04}
Smith, R.J. et al., 2004, MNRAS in press
\bibitem[\protect\citeauthoryear{Spergel et al.}{2003}]{spergel03}
Spergel, D.N., Verde, L., Peiris, H.V., Komatsu, E., Nolta, M.R.,
Bennett, C.L., Halpern, M., Hinshaw, G., Jarosik, N., Kogut, A.,
Limon, M., Meyer, S.S., Page, L., Tucker, G.S., Weiland, J.L.,
Wollack, E. \& Wright, E.L., 2003, ApJ Supplement Series, 148, 175
\bibitem[\protect\citeauthoryear{Webster}{1976}]{webster76}
Webster, A.S., 1976, MNRAS, 175, 61
\bibitem[\protect\citeauthoryear{Webster}{1982}]{webster82}
Webster, A.S., 1982, MNRAS, 199, 683
\bibitem[\protect\citeauthoryear{Williger et al.}{2002}]{williger02}
Williger, G.M., Campusano, L.E., Clowes, R.G. \& Graham, M.J.,
2002, ApJ, 578, 708

\end{thebibliography}
\end{document}